\documentclass[reprint,amsmath,amssymb,aps]{revtex4-1}
\usepackage{graphicx}
\usepackage{dcolumn}
\usepackage{bm}
\usepackage{float}
\usepackage{color}
\usepackage[dvipsnames]{xcolor}
\usepackage{hyperref}
\usepackage{graphicx}
\usepackage{braket}
\usepackage{appendix}
\usepackage{chemmacros}
\hypersetup{colorlinks=true, citecolor=blue, urlcolor=blue, linkcolor=blue}
\begin{document}

\preprint{APS/123-QED}

\title{Stabilization of A-site ordered perovskites and formation of spin-half antiferromagnetic lattice: CaCu$_3$Ti$_4$O$_{12}$ and CaCu$_3$Zr$_4$O$_{12}$}

\author{Jatin Kumar Bidika}
\email{Equal Contributors}
\author{Amit Chauhan}
\email{Equal Contributors}
\author{B. R. K. Nanda}
\email{nandab@iitm.ac.in}

\affiliation{
Condensed Matter Theory and Computational Lab, Department of Physics, Indian Institute of Technology Madras, Chennai - 600036, India\\
Center for Atomistic Modelling and Materials Design,
Indian Institute of Technology Madras, Chennai - 600036, India\\
Functional Oxide Research Group,
Indian Institute of Technology Madras, Chennai - 600036, India
}

\date{\today}
\begin{abstract}
A-site ordered perovskites, CaCu$_3$B$_4$O$_{12}$, which are derivatives of conventional ABO$_3$ perovskites, exhibit varying electronic and magnetic properties. With the objective of examining the role of Cu in this work, we have studied  CaCu$_3$Ti$_4$O$_{12}$ and CaCu$_3$Zr$_4$O$_{12}$ and presented the cause of the crystallization of A-site ordered perovskite from conventional ABO$_3$ perovskite and the underlying mechanism leading to the stabilization of non-trivial and experimentally estabilished G-type antiferromagnetic (G-AFM) ordering in these systems. The first-principles electronic structure calculations supplemented with phonon studies show that the formation of A-site ordered perovskite is driven by Jahn-Teller distortion of the CuO$_{12}$ icosahedron. The crystal orbital Hamiltonian population analysis and magnetic exchange interactions estimated using spin dimer analysis infers that the nearest and next-nearest-neighbor interactions (J$_1$ and J$_2$) are direct and weakly ferromagnetic whereas the third-neighbor interaction (J$_3$) is unusually strong and antiferromagnetic driven by indirect superexchange mechanism. The structural geometry reveals that stabilization of G-AFM requires J$_1$ $<$ 2J$_2$, J$_1$ $<$ 2J$_3$. The experimental and theoretical values of Neel Temperature agrees well for $U$ $\approx$ 7 eV, highlighting the role of strong correlation. The magnetic ordering is found to be robust against pressure and strain.
\end{abstract}

\maketitle
\section{\label{sec:level1}Introduction}
In past many decades perovskite transition metal oxides with formula ABO$_3$ have been extensively studied as these pristine compounds and their derivatives facilitate a plethora of intriguing electronic and magnetic properties both for fundamental studies and novel applications \cite{Interesting_magnetic_1, Interesting_magnetoresistance_2, Simple_Perovskite_A_Birds_Eye_View_RAMADASS1978231, Ravindran1999_Simple_Perovskite}. These properties can be tuned by applying external pressure or strain, electric and magnetic fields or by changing the chemical composition \cite{Tuning_Magnetic_Properties_New,Vailionis_2011, Varignon_2016}. Here A is either a rare earth element or alkaline earth metal and B is a transition metal atom.  \par
Another class of perovskites, known as quadruple perovskites or A-site ordered perovskites is less explored despite having equally intriguing electronic structure \cite{A_Site_Perovskite_Intriguing_Property, CCPtO_B_Site_Inactive}. This family of perovskites with chemical formula AA$^\prime_3$B$_4$O$_{12}$ are derived from the cubic perovskite structure in which A and A$^\prime$ site cations form an ordered structure with 1:3 composition. While A is an alkaline earth metal, A$^\prime$ belongs to a transition metal atom. The CaCu$_3$Ti$_4$O$_{12}$ (CCTiO) is an antiferromagnetic insulator and possess Neel temperature (T$_N$) of $\approx$ 25 K and a giant dielectric constant of the order $10^{4}$ over a wide range of temperature \cite{Subramanian2000, CCTO_2_exp_investigation,CCTO_3_exp_investigation,Neel_Temperature_TN, Neutron_diffraction_CCTO}. Furthermore, the x-ray absorption spectroscopy measurements confirm the antiferromagnetic ordering of CCTiO \cite{XAS_CCTO}. The manganate, CaCu$_3$Mn$_4$O$_{12}$, exhibits colossal magnetoresistance even under a lower magnetic field due to  Mn$^{4+}$ character ~\cite{CCMO_Zeng_4_1999}. The high temperature phase of CaCu$_3$Fe$_4$O$_{12}$ exhibits a ferrimagnetic ground state where Fe is in a homogeneous valence state (Fe$^{4+}$) whereas the low temperature phase shows charge disproportionation from Fe$^{4+}$ to Fe$^{5+}$ ~\cite{CCFO_HaO_5_2009}. The Zhang-Rice singlet state was observed in the ferromagnetic metal CaCu$_3$Co$_4$O$_{12}$  and ferromagnetic half metal CaCu$_3$Ni$_4$O$_{12}$  \cite{CCCO_Zhang_Rice_singlet_state_2013, CCNO_Zhang_Rice_Singlet_state}. The candidates of CCBO family with B-site as 4$d$ transition metal elements CaCu$_3$Ru$_4$O$_{12}$ and CaCu$_3$Rh$_4$O$_{12}$ exhibits Pauli paramagnetic metallic state \cite{CCRhO_Paramagnetic_Metal,Pauli_Paramagnetic_CCRO}. The CCRuO is a heavy-fermionic system where Cu and Ru possess 2+ and 4+ oxidation states, whereas in CCRhO, the unusual oxidation states $\approx$ 2.8 + and 3.4 + was attributed to the larger crystal-field splitting.\par
There are two issues that need to be addressed to understand and tailor the fundamental phenomena involving the couplings of spin, charge, and lattice degrees of freedom in CCBO. Firstly, the empirical observations suggest that the B-site hardly influences the crystallization in the quadruple perovskite structure and therefore the role of Cu in achieving the structural stability should be explained. The second issue involves the magnetic ordering when the B-site is magnetically inactive and composed of sp-elements. For example CaCu$_3$Ge$_4$O$_{12}$ and  CaCu$_3$Sn$_4$O$_{12}$ systems are found to be fully ferromagnetic \cite{Ferromagnetic_A_Site_Ordering}. However, if the B-site is magnetically inactive but composed of $d$-elements, as in the case of CCTiO, CCZrO and CaCu$_3$Pt$_4$O$_{12}$, the system stabilizes in an unnatural antiferromagnetic state \cite{Toyoda2013, CCPtO_B_Site_Inactive}. One of the theoretical studies on CCTiO/CCZrO by Toyoda \textit{et al.}, suggests that the nearest-neighbor Cu spins prefer weak ferromagnetic coupling while the third-neighbor Cu spins favour strong antiferromagnetic coupling in these spin-half lattices \cite{Toyoda2013}. The ferromagnetic coupling was explained through the indirect exchange interaction via Cu-O-Cu path whereas the strong antiferromagnetic coupling was explained through superexchange interaction via a zig-zag Cu-O-Ti-O-Cu path. Furthermore, the estimated value of Neel temperature was quite high ($\approx$ 70 K) as compared to the experimentally observed value of 25 K. Contrary to Toyoda \textit{et al.}, Lacroix \cite{Lacroix_1980} considered the exchange paths that exclude the oxygen atoms (Cu-Ti-Cu interaction) to explain the ground state magnetic configuration. Therefore, the mechanism of magnetic couplings is still unsettled. \par


To address the aforementioned issues, we have investigated the experimentally synthesized CCTiO \cite{Collosal_Dielectric_Constant_CCTiO, Integrated_Local_Electronic_Study_CCTiO} and yet to be synthesized CCZrO with the aid of density functional theory (DFT) calculations. The d$^0$ configuration of Ti and Zr provides a Cu only spin lattice and make the Fermi level devoid of any B-site electrons. This implies that the Cu is responsible both for the magnetization as well as the symmetry lowered crystallization of these compounds. Furthermore, the present study brings insight on the effect of strong correlation on the magnetic structure of these systems which is missing in the literature. Also, in correlated oxide systems, often strain and pressure are employed to induce a magnetic transition. In this work, we have examined whether such a transition can be achieved here as well. To establish the mechanism of magnetic exchange interactions, the electron hopping integrals involving the orbitals of various atom pairs (Cu-O, Cu-Ti, Cu-Cu etc.) in different exchange paths are analyzed using projected crystal orbital Hamilton population (pCOHP). The strength of exchange interactions are obtained by employing spin dimer analysis.\par
From electronic structure analysis we find that the structural transition from a cubic to  A-site ordered structure is driven by Jahn-Teller distortion of the Cu-O complex. The distortion is necessary to lower the  density of degenerate states at the Fermi level. The analysis of exchange interaction infers that the nearest and next-nearest neighbor (J$_1$ and J$_2$) interactions are direct and hence weakly ferromagnetic. However, the third-neighbor (J$_3$) interaction is found to be relatively stronger and antiferromagnetic, driven by long range superexchange mechanism leading to the formation of G-type antiferromagnetic ordering. The exchange pathways are validated through COHP analysis. The exact match between experimentally observed value of T$_N$ ($\approx$ 25 K) is found for $U$ $\approx$ 7 eV, signifying the importance of electron correlation in this family of compounds. The revised exchange interaction mechanisms in this present study will pave the way to reexamine the electronic and magnetic structure of other members of this A-site ordered perovskite family. 

\section{Structural and computational details}
The regular perovskite (ABO$_3$) structure is shown in Fig. \ref{Fig1}(a). Had the CCBO stabilized in this structure Ca and Cu would have shared the A site. The stable A-site ordered perovskite configuration is shown in Fig. \ref{Fig1}(b). It contains two formula units per unit cell. The B-site transition metal elements form BO$_6$ distorted octahedral complexes whereas the A site doped Cu atoms form  square planar CuO$_4$ complexes in the $xy$, $yz$ and $xz$ planes. Each distorted octahedral complex is interlocked with the six square planar complexes. The distortion of the regular perovskite leading to the A-site ordered perovskite structure is discussed later.

The first-principles based electronic structure calculations are performed using the full-potential linearized augmented plane wave (FP-LAPW) \cite{FPLAPW_1979} method as implemented in the WIEN2k \cite{WIEN2k} code. For all calculations, the generalized gradient approximation (GGA) was considered for the exchange-correlation functional. A $8 \times 8 \times 8$  k-mesh was used for the Brillouin Zone integration and is found to be sufficient to estimate the electronic structure with reasonable accuracy.
The mean-field based  parametric Hubbard $U$ formalism is adopted to examine the effect of on-site Coulomb repulsion giving rise to strong correlation. It is done through the rotationally invariant Dudarev approach with $U_{eff}$ = $U-J$ \cite{GGA_U_Analysis, GGA_U_1991}  where $U$ and $J$ are Hubbard and Hund's exchange parameters respectively. Herein, the value of J is considered as zero. Therefore, the $U_{eff}$ becomes $U$. The convergence criterion for the total energy and charge density was set as 10$^{-4}$ Ry.
The electronic structure calculations were performed using both experimental and optimized lattice parameters for CCTiO and optimized lattice parameter for CCZrO. First we performed the spin-polarized calculations within GGA, and then considered the Coulomb interaction through GGA+$U$ calculations. The density functional perturbation theory (DFPT) \cite{Giannozzi1991,Gonze1997} approach is adopted for the phonon calculations without considering the $U$. The force constants obtained from the DFPT method is taken into account through the phonopy code as implemented in Vienna ab-initio Simulation Package (VASP) ~\cite{Togo2010,Togo2008}. For the phonon calculations a $1 \times 1 \times 2$ supercell is considered. 

To understand the nature of mechanism for the exchange paths, the pCOHP is calculated for different atom pairs. The pCOHP is a partitioning of the
band-structure energy in terms of orbital-pair contributions. In principle, the pCOHP indicates bonding and anti-bonding energy regions for a specified energy range. The pCOHP is expressed as:
\begin{equation}
    pCOHP_{\mu,\nu} (E,k) = \sum_j R [P_{\mu \nu j}^{proj}(k)H_{\nu \mu }^{proj} (k)] \times \delta (\epsilon_j(k) - E) 
\end{equation}
where, $P_{\mu \nu j}^{proj}(k)$ is the projected density matrix for band j at k-point and $H_{\nu \mu}$ is the hybridization strength of orbital pairs $\mu$ and $\nu$ centered at two atoms. The pCOHP calculations were performed by using the Local-Orbital Basis Suit Towards Electronic-Structure Reconstruction (LOBSTER) code \cite{Dronskowski1993,Deringer2011}.

The optimised lattice parameters, the B-O and Cu-O bond lengths, the local and global O-B-O, O-Cu-O, B-O-B and Cu-O-B bond angles, along with the experimental lattice parameters \cite{Subramanian2000, Yamada2010} of the CCTiO and CCZrO are given in Table-I.
\begin{table*}
    \setlength{\arrayrulewidth}{0.07mm}
    \setlength{\tabcolsep}{8pt}
    \renewcommand{\arraystretch}{1.4}
    \centering
    \caption{The B site transition metal element in CCBO system,  optimized and experimental lattice parameter \textit{a} in ({\AA}), the M-O, Cu-O bond lengths in ({\AA}), O-B-O, O-Cu-O, B-O-B and Cu-O-B bond angles are listed for CCBO system.}
    \begin{tabular}{cccccccc} \hline \hline
      CCBO & \textit{a} & B-O & Cu-O & O-Cu-O & O-B-O & B-O-B & Cu-O-B    \\ \hline
      CCTiO (optimized)  & 7.448 & 1.98 &  1.96 & 96.34$^{\circ}$, 83.66$^{\circ}$ & 
         89.02$^{\circ}$, 90.98$^{\circ}$ &  139.89$^{\circ}$ & 109.57$^{\circ}$     \\
    CCTiO (experiment)  & 7.391 & 1.97 &  1.94 & 90.69$^{\circ}$, 89.31$^{\circ}$ & 
         87.09$^{\circ}$, 92.91$^{\circ}$ &  139.94$^{\circ}$ & 110.02$^{\circ}$     \\
         
      CCZrO (optimized)  & 7.887 & 2.13 &  2.00 & 81.33$^{\circ}$, 98.66$^{\circ}$ & 
         88.57$^{\circ}$, 91.42$^{\circ}$ &  135.22$^{\circ}$ & 111.58$^{\circ}$     \\ \hline\hline 
    \end{tabular}
    \label{Crystalographic data of the series of CCBO systems.}
\end{table*}
\section{Results and Discussion}
\subsection{Formation of the A-site ordered perovskite structure: Role of Cu-$d$ electrons}
\begin{figure}
\includegraphics[angle=-0.0,origin=c,height=10cm,width=8cm]{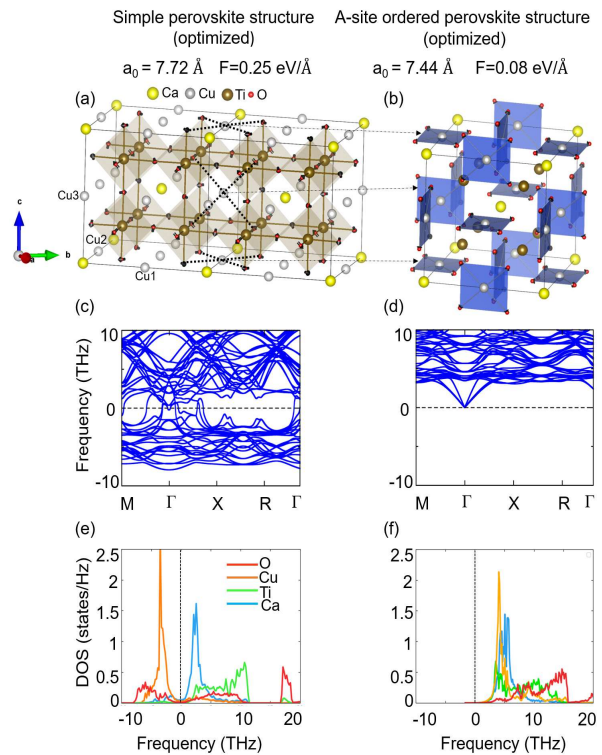}
\caption{(a) The hypothetical volume-optimized structure of CCTiO modelled based on single Perovskite structure (ABO$_3$). In this eight formula unit cell, six of the A (Ca) atoms are replaced by Cu atoms to form the stoichiometry CaCu$_3$Ti$_4$O$_{12}$. The octahedra (OH) and square planar (SP) complexes are shown. The force \textcolor{blue}{(F)} at the atomic sites are mapped to give an insight on the possible displacement of anions leading to intra- and inter- distortion of the OH and SP complexes. (b) The stable crystal structure as obtained from experiment. (c, d) and (e, f) respectively display the phonon band structures and atom resolved density of states corresponding to the cubic and experimental structure.}
\label{Fig1}
\end{figure}

In the conventional single perovskite oxides ABO$_3$, B forms a BO$_6$ octahedron while A forms an AO$_{12}$ icosahedron. This structure is stable when A is an electron donor from the alkali/alkaline or rare earth family and does not participate in covalent bonding to form the band structure. However, when a transition metal element like Cu occupies the A sites, as shown in Fig. \ref{Fig1}(a), the phonon band structures yield large number of imaginary frequencies (see Fig. \ref{Fig1}(c)) indicating huge instability for the simple perovskite structure of CaCu$_3$Ti$_4$O$_{12}$. This hypothetical perovskite structure is built by considering a eight formula unit CaTiO$_3$ cell and replacing six of the Ca by Cu. While there are many possible arrangements of Ca and Cu that can be thought of in this hypothetical structure, they are broadly put together into two groups. One where the Ca cations are close to each other and in the other case they are away from each other. Our total energy calculations suggest that the former is unstable compared to the later by $\sim$ 3.3 eV. Therefore, the minimum energy configuration is built by keeping the Ca ions away from each other as shown in Fig. 1(a).

Now we shall further analyze the phonon spectra to establish the stabilization of the A-site ordered perovskite structure out of the simple ABO$_3$ structure. In Fig. \ref{Fig1}(e), we have plotted the site projected phonon DOS for the latter and it shows that the dominant contribution to these imaginary frequencies comes from the Cu and O sites, whereas the Ca sites have very nominal contribution which infers the negligible role of Ca towards structural instability. The presence of dominant imaginary frequencies is attributed to the large electrostatic force exerted among the Cu cation and O-anions in the Cu-O icosahedron. As a consequence the system undergoes a symmetry lowering structural transition. The force driving this transition are indicated in Fig. \ref{Fig1}(a) and the resultant displacement leading to formation of square-planar complexes are indicated using dashed lines. The ionic displacements by the lighter O atoms lead to the formation of A-site ordered structure (Fig. \ref{Fig1}(b)). The dynamical stability of this structure is inferred from the absence of imaginary frequencies (Fig. \ref{Fig1}(d), (f)).

\begin{figure*}
\includegraphics[angle=-0.0,origin=c,height=9.5cm,width=18cm]{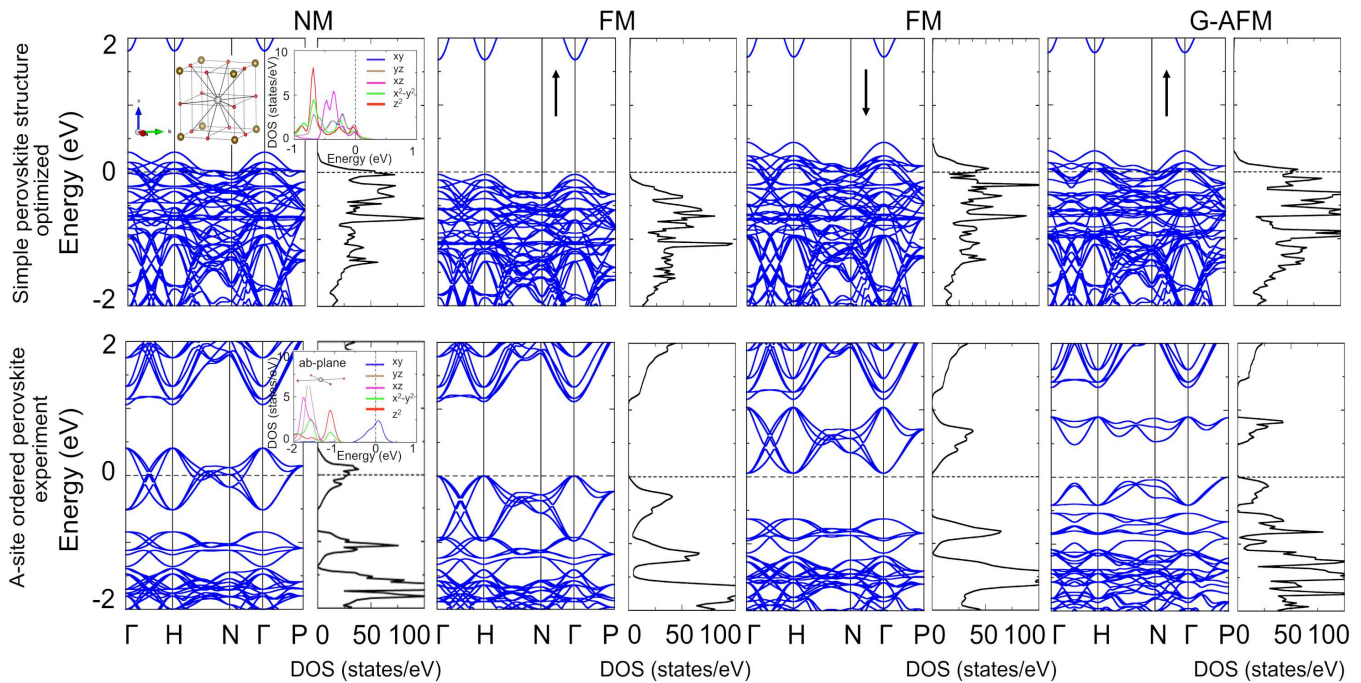}
\caption{Electronic structure, as represented through bands and DOS, for nonmagnetic, ferromagnetic and G-type antiferromagnetic states for optimized simple perovskite structure (top row), and experimental A-site ordered perovskite (bottom row). For the NM simple perovskite structure, even though the Cu1-$d$ states experience an icosahedron crystal field, the xz DOS differs from the xy and yz DOS. This is due to the fact that the Cu1-O-Cu2 and Cu1-O-Cu3 chains are periodic in the $ab$ and $bc$ planes which allows a continuous $d-p-d$ electron hopping. However, in the $ac-$ plane this hopping path is broken due to the presence of the Cu1-O-Ca-O-Cu1 chain and as a result the xz DOS differs from the degenerate xy and yz DOS. Similar inferences can be made for Cu2-yz and Cu3-xy orbitals.}
\label{Fig2}
\end{figure*}
The primary understanding towards the stabilization of the A-site ordered perovskite structure out of a simple cubic perovskite structure can be obtained by examining the electronic structure of the system. According to the Jahn-Teller theorem \cite{Jhan_Teller_Theorem1937}, if there are degenerate states occupying the Fermi level (E$_F$) the cation-anion complexes distort to reduces the covalent interaction and in turn to deplete the degenerate density of states (DOS) at the E$_F$. This results in reduction of kinetic energy and stabilization of the structure. Interestingly, depletion of DOS at E$_F$ can also occur through spin polarization which is well understood through the Stoner criterion. Through Fig. \ref{Fig2}, where the band structure and DOS are plotted, we examine the role of symmetry breaking on the electronic structure and magnetization of CCTiO. The top and bottom rows represent the case of simple perovskite and A-site order perovskite structure (experimental) respectively. The results for the nonmagnetic (NM), ferromagnetic (FM) and G-type antiferomagnetic (G-AFM) states are presented column wise. The non-magnetic configuration in the simple perovskite structure shows a large DOS at E$_F$ with a large number of degenerate bands. Since Ti 4+ charge state leads to a $d^0$ electronic configuration, these degenerate bands solely arise from the Cu-$d$ states. \par 

The Cu at A-site creates an icosahedron which can be viewed as three superimposing square-planar complexes lying on the xy, yz, and xz planes (see inset in top row of Fig. \ref{Fig2}). Therefore, the net effective crystal field is the sum of the three independent square planar crystal fields, and as a result the otherwise non-degenerate crystal field split orbitals overlap to form a new set of degenerate states occupying the E$_F$. As a consequence, the states in the vicinity of E$_F$ are highly degenerate. Since the conventional unit cell consists of six equivalent Cu atoms, there are thirty spin degenerate $d$-orbitals and the partial DOS indicates that eighteen of them cross the E$_F$. These degenerate states bring Jahn-Teller instability and therefore, the lattice undergoes distortion to give rise to isolated square planar complexes as discussed in the previous paragraph.  In the square-planar crystal field, the five-fold degenerate $d$-orbitals split into four manifolds with three non-degenerate and one two-fold degenerate states as demonstrated in the inset of bottom row of Fig. \ref{Fig2}. From each complex only one non-degenerate $d$-state crosses the E$_F$ and hence, altogether there are only six bands instead of eighteen. A filled shell electronic structure of Ca$^{2+}$ ion does not contribute to the JT distortion.\par

While the Jahn-Teller mediated lattice distortion depleted the DOS at E$_F$ by one-third, it is large enough not to stabilize the system in a non-magnetic ground state. However, the depletion of DOS at E$_F$ through the spin polarization opens up a gap at E$_F$ by pushing the spin-up bands below and spin-down bands above in energy as shown in Fig. \ref{Fig2}. With one unoccupied spin minority band for each Cu the system gives rise to a spin-1/2 lattice. From our total energy calculations on several spin arrangements of this spin-1/2 lattice  we find that the G-type aniferromagnetic ordering forms the ground state. The partial DOS for the latter is depicted in Fig. \ref{Fig5} and will be discussed later. Further understanding on the magnetization of this compound is obtained by calculating the magnetic exchange interactions which is discussed next.
\subsection{Exchange interactions for $d^0$ system}

\begin{figure}
\includegraphics[angle=-0.0,origin=c,height=7.5cm,width=8.5cm]{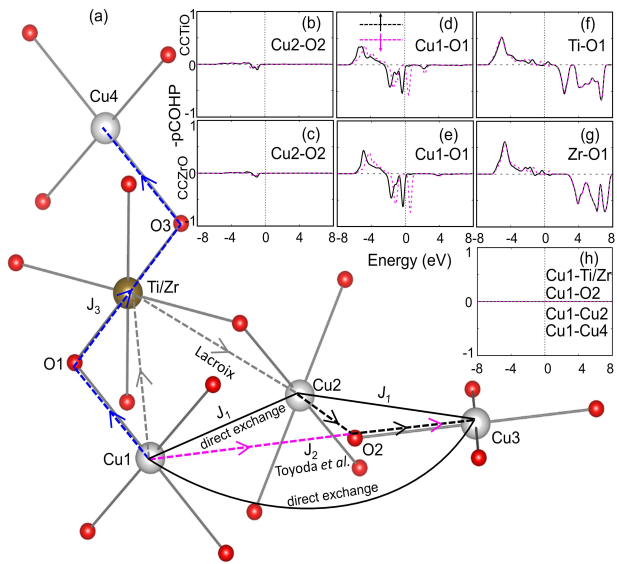}
\caption{(a) Depiction of exchange interaction paths for J$_1$, J$_2$ and J$_3$.(b-h) The pCOHP of various atom pairs in CCTiO and CCZrO. The negative (positive) values of -pCOHP indicates the antibonding (bonding) states respectively. }
\label{Fig3}
\end{figure}

\begin{figure}
    \centering
    \includegraphics[angle=-0.0,origin=c,height=13cm,width=8.6cm]{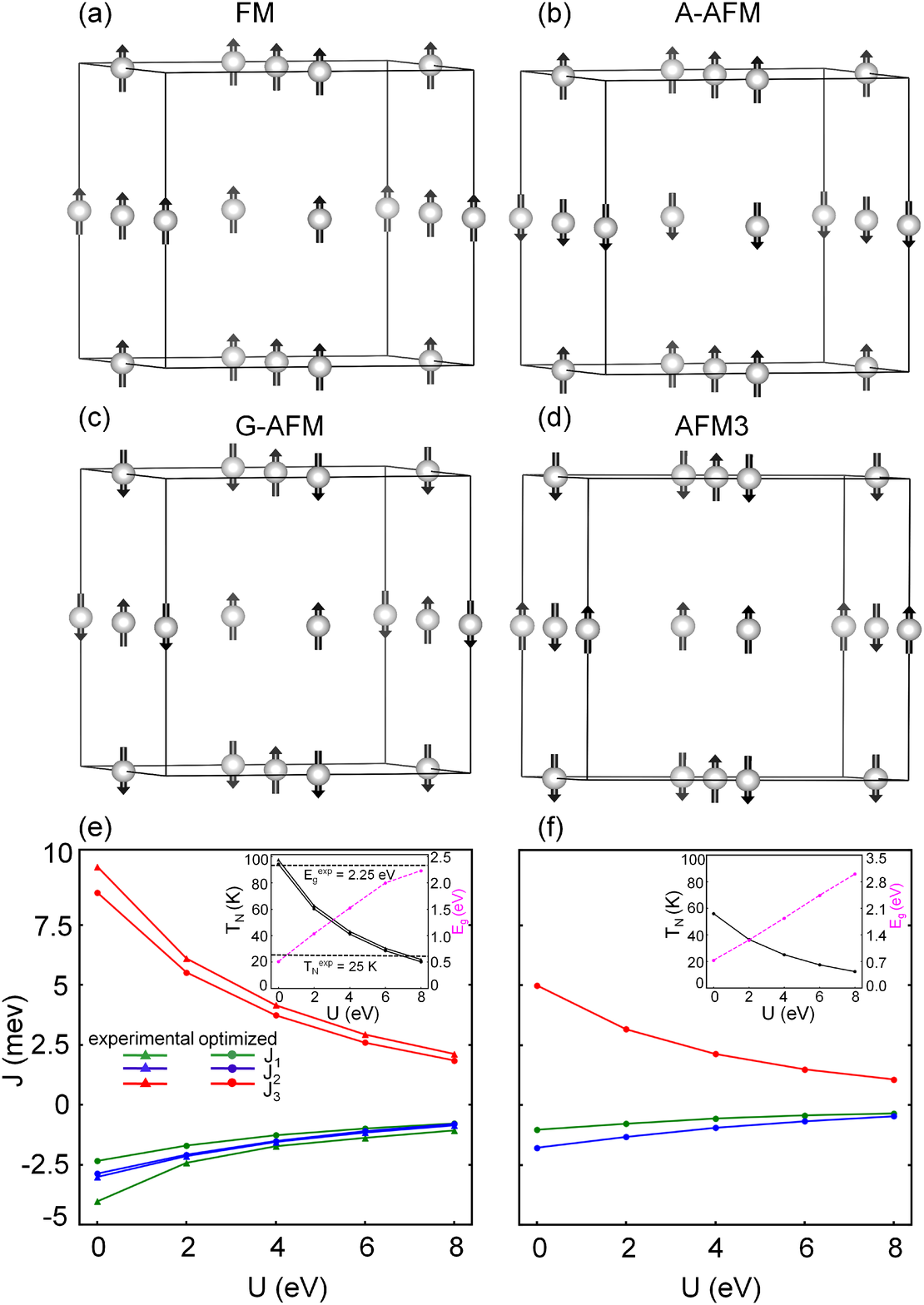}
    \caption{(a-d) The magnetic configurations, namely, FM, A-AFM, G-AFM and AFM-3. (e-f) The variation of exchange couplings J$_1$, J$_2$, J$_3$, and Neel temperature T$_N$ with $U$ (see insert) for experimental and optimized structures of CCTiO and CCZrO. The dashed lines represent the experimentally obtained values of T$_N$ and E$_g$ respectively.}
    \label{Fig4}
\end{figure}

To find the underlying mechanism governing the magnetic ordering in this system, a detailed study of spin-exchange interactions is necessary. In one of the works by Toyoda \textit{et al.} \cite{Toyoda2013}, the  exchange interactions  J$_1$, J$_2$ and J$_3$ (see Fig. \ref{Fig3}) were explained with the aid of a classical Heisenberg Hamiltonian, cluster model and charge density analysis. The J$_1$  and J$_2$ are found to be weakly ferromagnetic and indirect, mediated by superexchange mechanism via the Cu-O-Cu path as depicted through Fig. \ref{Fig3}(a). Contrary to J$_1$ and J$_2$, the J$_3$ was found to be strong and antiferromagnetic (J$_3$ $\approx$ 5 J$_1$ and J$_3$ $\approx$ 5 J$_2$), mediated by long-range superexchange interaction via Cu-O-Ti-O-Cu path. Even though the nature of magnetic couplings was similar for both CCTiO and CCZrO, the cause for the higher strength of J$_3$ in the former has not been discussed in this work. Furthermore, the estimated value of T$_N$ ($\approx$ 70 K) for CCTiO is nearly three times higher then the experimentally reported value  of ($\approx$ 25 K) \cite{ Neel_Temperature_TN}. 

In another study on CCTiO by Lacroix, perturbative Hamiltonian approach was adopted to propose that the exchange paths are defined through Cu-Ti-Cu interactions (see Fig. \ref{Fig3}(a)) \cite{Lacroix_1980}. The G-type magnetic structure was interpreted by considering the competition between superexchange interaction and spin anisotropy. Though the absolute value of the exchange interaction strengths were not determined, the J$_3$  was estimated to be ten times higher than J$_1$ and J$_2$. 

The aforementioned discussion suggests that an intricate analysis of the magnetic coupling is needed to determine the appropriate mechanism that stabilizes the G-type magnetic ordering as well as to accurately estimate the T$_N$ for this system. For this purpose we have employed the Noodelmann’s broken-symmetry spin dimer method \cite{Noodlemann's_method} on CCTiO and CCZrO. In this method the energy difference between high spin (HS) and broken symmetry (BS) configurations is given by
\begin{equation}
E_{HS} - E_{BS} = \frac{1}{2} S_{max}^2 J
\end{equation}
where J is related to the spin-dimer Hamiltonian $H$ = $\sum_{i<j} J_{ij} S_i \cdot S_j$ and $S_{max}$ is the maximum number of unoccupied spins of the dimer. For the present case each monomer has one unoccupied spin, hence, the above equation can be expressed as 
\begin{equation}
E_{HS} - E_{BS} = \frac{J}{2} 
\end{equation}
The E$_{HS}$ and E$_{BS}$ are evaluated by performing the DFT calculations for HS and BS configurations, respectively. To evaluate the J$_i$'s strength, four different magnetic configurations, namely, FM, A-type, G-type and AFM3 are considered and are shown in Fig. \ref{Fig4}(a)- Fig. \ref{Fig4}(d). Using the spin dimer Hamiltonian for these magnetic configurations, we have evaluated the spin-exchange energies (per f. u.) in terms of the exchange parameters and are expressed as : \par
\begin{eqnarray}
E_{FM} = \frac{1}{8}(12J_1 + 24J_2 + 24J_3),\notag\\
E_{AAFM} = \frac{1}{8}(4J_1 - 8J_2 - 24J_3),\notag\\
E_{GAFM} = \frac{1}{8}(-12J_1 + 24J_2 - 24J_3)\notag,\\ 
E_{AFM3} = \frac{1}{8}(-4J_1 - 8J_2 + 8J_3).
\end{eqnarray}

These spin-exchange energies are mapped to the total energy obtained from the DFT calculations for the respective magnetic configuration. Further, using Eq. 2, the $J$ values are estimated as a function of $U$ for both experimental and optimized crystal structures of CCTiO. For CCZrO the exchange parameters are calculated only for the optimized structure as it is yet to be experimentally synthesized. Further, the Neel temperature was estimated using Eq. 5,  where $\Theta_{CW}$, K$_B$, Z$_i$ and $\mu$ represents the Curie-Weiss temperature, Boltzmann constant, coordination number corresponding to each J$_i$ interaction, and the mean field constant respectively, and the results are shown in Fig. \ref{Fig4}. For CCTiO, we have considered the $\mu$ = $\frac{\theta_{cw}}{T_N}$ (= 1.30) as reported in one of the experimental work \cite{A_Site_Perovskite_Intriguing_Property}. However, the CCZrO is yet to be experimentally synthesized. Therefore, for CCZrO, the $\mu$ value is taken as that of CCTiO. This value of $\mu$ is fair enough to consider as both former and latter adopts the same structural framework and both Zr and Ti have d$^0$ electronic configuration. Here, the positive (negative) values of $J$ indicate AFM (FM) coupling between neighboring spins. As expected the J$_1$ and J$_2$ are weak and ferromagnetically coupled whereas the J$_3$ is antiferromagnetic. Keeping these modes of interactions intact, we found the conditions to stabilize different magnetic ordering in these systems and are listed in Table-II.
\begin{table}
    \centering
    \caption{The conditions for the stabilization of G-AFM, FM, A-AFM and AFM3 magnetic configurations.}
    
\begin{tabular}{cc}\hline \hline
    Magnetic& Conditions \\
    Configurations & (in absolute values of J$_i$) \\  \hline
        G-AFM & J$_1$ $<$ 2J$_2$, \hspace{0.2cm} J$_1$ $<$ 2J$_3$ \\
        FM & J$_1$ $>$ 2J$_3$, \hspace{0.2cm} J$_2$ $>$ J$_3$ \\
        A-AFM & J$_1$ $>$ 2J$_2$, \hspace{0.2cm} J$_1$ $>$ -4J$_3$ \\ 
        AFM3 & J$_1$ $<$ -4J$_3$, \hspace{0.2cm} J$_2$ $<$ J$_3$ \\ 
        \hline\hline
    \end{tabular} 
     \label{Table-II}
\end{table}
To identify the correct spin exchange pathways, the bonding strength between various atom pairs is calculated using COHP and the results are shown in Fig. \ref{Fig3}(b)-Fig. \ref{Fig3}(h). The J$_1$ and J$_2$ paths adopted by Toyoda \textit{et al.} includes intermediate oxygen atoms through Cu2-O2-Cu3 pd$\pi$ and Cu1-O2-Cu3 pd$\sigma$ covalent interactions respectively (see Fig. \ref{Fig3}(a)). However, the COHP calculations suggest that for Cu1-O2 and Cu2-O2 atom pairs hardly have any covalent bonding. Hence, such indirect exchange pathways are least probable. As discussed earlier, Lacroix consider Cu-Ti-Cu path for both J$_1$ and J$_2$ and explained the strength through Cu-Ti electron transfer integral. However, the COHP result shown in Fig. \ref{Fig3}(h) predict the Cu1-Ti interaction to be zero and hence such indirect exchange pathways are improbable. Furthermore, the nearest and next-nearest CuO$_4$ square planer complexes and thereby the Cu-$d$ orbitals, occupying the unpaired electrons, are orthogonal to each other. Therefore, the overlapping of these orbitals in the nearest and next-nearest neighborhood is negligible. This implies that the J$_1$ and J$_2$ are mediated through direct exchange mechanism. Similar inference are obtained for CCZrO.

The third neighbor CuO$_4$ complexes are non-orthogonal. However, due to large separation between Cu1 and Cu4 ions, the hybridization between Cu-$d$ orbitals is negligible. Since the direct exchange operates for short inter-nuclear separations, hence J$_3$ can only be explained through indirect exchange mechanisms. The COHP for Cu1-O1 and Ti/Zr-O1 pairs, plotted in Fig. \ref{Fig3}(d)-Fig. \ref{Fig3}(g), show stronger overlap among the orbials of these pairs. Furthermore, identical position for bonding and antibonding states in the energy range for both of them favours the indirect exchange path of Cu1-O1-(Ti/Zr)-O3-Cu4. This is in agreement with Tyoda \textit{et al.} who have predicted indirect superexchange mediated antiferromaggetic interaction along this path.

\begin{equation}
    \Theta_{CW} = - \frac{S(S+1)}{3K_B} \sum_i Z_i J_i, \hspace{0.5cm} T_N = \frac{\mid \Theta_{CW} \mid}{\mu} 
\end{equation}

The relative strength of J$_1$, J$_2$ and J$_3$ as obtained in this mean-feild study within the framework of GGA implies the stabilization of the G-type ordering. However, the T$_N$ estimated using Eq. 4 is found to be $\approx$ 95K, which is approximately four times the experimentally observed T$_N$ value which is an expected overestimation in the mean-field framework. In the coming section we will see that this can be addressed by taking into account the electron correlation effect.

\section {Role of onsite Coulomb repulsion and lattice distortions on magnetism}
\begin{figure*}
    \centering
    \includegraphics[angle=-0.0,origin=c,height=12.8cm,width=18cm]{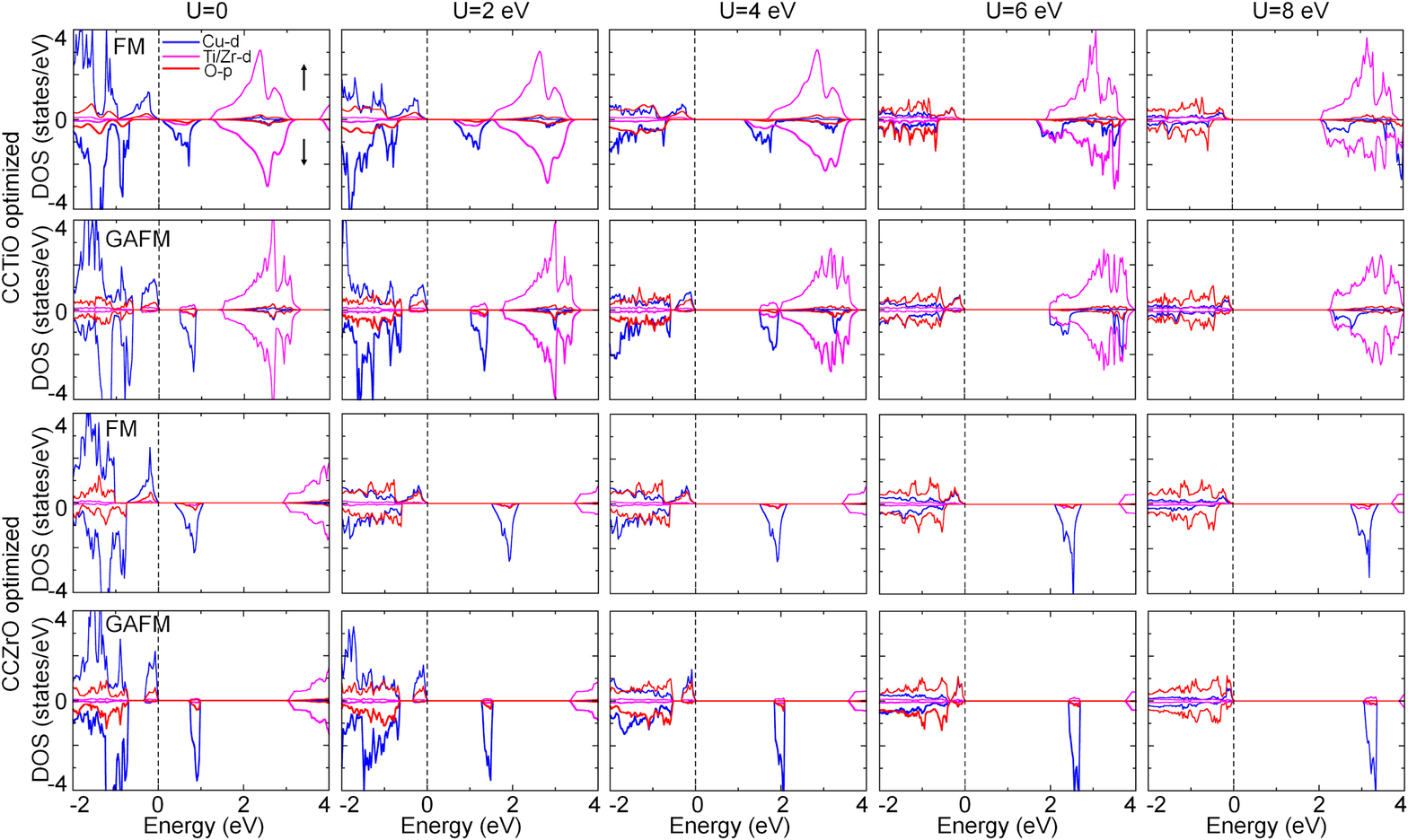}
    \caption{The projected density of states of FM and G-AFM states as a function of $U$ for optimized CCTiO (first and second row) and CCZrO (third and fourth row). The up and down arrows represent the spin-up and spin-down channels. The Fermi energy is set at zero.}
    \label{Fig5}
\end{figure*}

The effect of strong correlation on the electronic structure of CCTiO and CCZrO are analyzed through DFT+$U$ calculations. In Fig. \ref{Fig5} we have plotted the orbital and spin resolved DOS for CCTiO and CCZrO as a function of $U$. As Ti/Zr exhibits $d^0$ electronic configuration, the Ti/Zr $d$ states reside in the conduction band and are inactive. For CCZrO, the Zr-$d$ states are located far above the E$_F$ relative to the CCTiO. As already been discussed the system exhibits a band gap through magnetization even in the absence of on-site Coulomb repulsion. Increasing the strength of $U$ localizes the Cu-$d$ states and pushes the valence and conduction band apart to widen the band gap. The localization reduces the strength of covalent interactions and therefore it  influences the indirect exchange interaction significantly. 
As can be seen from Figs. \ref{Fig4}(e) and \ref{Fig4}(f), the strength of J$_1$, J$_2$, and J$_3$ are plotted as a function of $U$ for CCTiO and CCZrO respectively. While J$_3$ decreases rapidly, J$_1$ and J$_2$ remain almost unchanged. Qualitatively one can express the magnetic exchange interaction between two states $\ket{m}$ and $\ket{m\prime}$ as $ J_{eff}$ = $f(t{_{mm\prime}}, U$) - $\bra{m}V_{12}\ket{m^\prime}_{exch}$, where $t_{mm\prime}$ is the effective hopping integral (direct or indirect). The function $f$ depends on the number of intermediate paths connecting m and m$\prime$. The second term is the direct exchange interaction. For J$_1$ and J$_2$, the first term vanishes due to the orthogonality of the Cu-$d$ orbitals and hence only the second term contributes which is independent of the onsite Coulomb repulsion as also inferred from Figs. \ref{Fig4}(e) and \ref{Fig4}(f). For J$_3$, the first term dominates and as a result J$_3$ decays rapidly with increasing onsite Coulomb repulsion $U$. This further validates the fact that J$_1$ and J$_2$ are direct exchange interactions while J$_3$ is indirect.

Figs. \ref{Fig4}(e) and \ref{Fig4}(f) enable us to calculate T$_N$ as a function of $U$ for CCTiO and CCZrO respectively and the results are shown in the insets. For CCTiO, the theoretical estimation of T$_N$ and bandgap (E$_g$) matches with the experimental one for $U$ $\approx$ 7 and 8 eV respectively \cite{HUANG2017}. Therefore, the strength of $U$ lie in the range of 7-8 eV, signifying that these systems are strongly correlated. For $U$ = 7 eV we predict the T$_N$ of CCZrO to be $\approx$ 15 K. In the case of CCTiO we find that the exchange interaction strengths for the optimized structure vary little with that of the experimental structure (see Fig. \ref{Fig4}(e)). Therefore, the predicted T$_N$ is expected to match with the experimental value for yet to be synthesized CCZrO.
\begin{table}
    \centering
     \caption{The T$_N$ (in K) of optimised CCTiO and CCZrO corresponding to 2 \% uniform compression, expansion and 2 \% epitaxial tensile strain with $U$ as 7 eV. The T$_N$ of equilibrium structure is also provided for comparison.}

\begin{tabular}{ccc} \hline \hline
        External stimuli & CCTiO & CCZrO  \\ \hline
        Uniform Compression & 29.75 & 17.16 \\
        \textbf{Equlibrium Structure} & \textbf{25} & \textbf{15} \\ 
        Uniform Expansion & 19.41 & 11.99 \\
        Epitaxial Tensile Strain & 23.10 & 14.13 \\
        \hline\hline
    \end{tabular}
   \label{Uniform pressure and strain effect on T$_N$.}
\end{table}
Often strain and pressure are used  to tune the magnetic properties of correlated oxides in general and perovskites in particular \cite{Nanda_2009, Marthinsen_2016, Nanda_2010, Meng2873}. To examine if these are critical factors in the case of CCTiO and CCZrO, we estimated the J$_i$s and hence the T$_N$ as a function of strain and pressure. While the G-AFM ordering remains robust within $\pm$ 2\% strain and pressure the T$_N$ was found to be tunable by 10 and 5K for CCTiO and CCZrO respectively (see Table-III).

\section{Summary}
To summarize, with the aid of density functional calculations, phonon studies, and spin dimer analysis we have examined the structural stability, electronic and magnetic structures of A-site ordered perovskites CCTiO and CCZrO. We show that the symmetry lowered transition from regular perovskite structure to A-site ordered perovskite structure is driven by Jahn-Teller distortion, where CuO$_{12}$ icosahedron gives rise to a CuO$_4$ square-planar complex. The underlying mechanism that  stabilizes the experimentally observed G-type antiferromagnetic ordering of the Cu spin-1/2 lattice is explained by calculating nearest, next-nearest and third-nearest exchange paths. By analyzing the crystal orbital Hamiltonian population we established that the nearest and next-nearest exchange interactions are direct and weakly ferromagnetic and the third-neighbor is indirect and strongly antiferromagnetic. The stabilization of G-AFM ordering is due to the fact the strength of J$_3$ is grater than half of J$_1$. This addresses the ambiguities on the nature of exchange interactions reported in the literature. The correlation effect has a major role in determining the Neel temperature. By pursuing DFT+$U$ calculations we show that the experimental and theoretical values of T$_N$ agrees well for $U$ $\approx$ 7 eV. Furthermore, we show that the magnetization and spin arrangements are robust against pressure and strain.

As a whole we believe that the present study will pave the way to revisit the electronic and magnetic structure of other members of the A-site ordered perovskite family.

\section{Acknowledgement}
The authors would like to thank HPCE, IIT Madras for providing the computational facility. BRKN would like to thank Debakanta Samal for fruitful discussions. This work is funded by the Department of Science and Technology, India, through grant No. CRG/2020/004330.

\end{document}